# Propagation of a magnetic domain wall in magnetic wires with asymmetric notches

A. Himeno, T. Okuno, S. Kasai, and T. Ono

*Institute for Chemical Research, Kyoto University, Uji, Kyoto 611-0011, Japan*

S. Nasu

*Graduate School of Engineering Science, Osaka University, Toyonaka, Osaka 560-8531, Japan*

K. Mibu

*Research Center for Low Temperature and Materials Sciences, Kyoto University, Uji, Kyoto 611-0011, Japan*

T. Shinjo

*International Institute for Advanced Studies, Soraku-gun, Kyoto 619-0225, Japan*

**Abstract**

The propagation of a magnetic domain wall (DW) in a submicron magnetic wire consisting of a magnetic/nonmagnetic/magnetic trilayered structure with asymmetric notches was investigated by utilizing the giant magnetoresistance effect. The propagation direction of a DW was controlled by a pulsed local magnetic field, which nucleates the DW at one of the two ends of the wire. It was found that the depinning field of the DW from the notch depends on the propagation direction of the DW.

**Corresponding author:** Atsushi Himeno

Institute for Chemical Research, Kyoto University, Uji, Kyoto 611-0011, Japan

*e*-mail: himeno@ssc1.kuicr.kyoto-u.ac.jp

Tel: +81-774-38-3105

Fax: +81-774-38-3109

The magnetic domain structure and the magnetization reversal process in magnetic nanostructures are controllable by modifying the sample shape. In a magnetic wire with submicron-width, the magnetization is restricted to be parallel to the wire axis due to the magnetic shape anisotropy, and the magnetization reversal takes place by nucleation and propagation of a magnetic domain wall (DW). Therefore, the magnetic wire is a simple model system to investigate the propagation of a magnetic DW. Several attempts have been reported to control nucleation and propagation of a DW.[1-5]

In this letter, we present a study on the propagation of a magnetic DW in a submicron magnetic wire with asymmetric notches. The propagation of the DW was detected by utilizing the giant magnetoresistance effect[2]. We show that the depinning field of the DW from the asymmetric notch depends on the propagation direction of the DW. We also show that the DW moves notch by notch by applying a pulsed local magnetic field.

The samples were fabricated onto thermally oxidized Si substrates by means of *e*-beam lithography and lift-off method. Figure 1 shows a schematic illustration of a top view of the whole sample. A magnetic wire has a trilayered structure consisting of $Ni_{81}Fe_{19}$(5 nm)/Cu(20 nm)/$Ni_{81}Fe_{19}$(20 nm). The main body of the magnetic wire has four notches with asymmetric shape. The sample has four current-voltage probes made of nonmagnetic material, Cu. Further, it has two narrow Cu wires crossing the ends of the magnetic wire and a wide Cu wire covering the notched part of the magnetic wire. These Cu wires are electrically insulated from the magnetic wire by SiO$_2$ layers of 50 nm in thickness. A flow of a pulsed electric current in each Cu wire can generate local magnetic fields, $H_L$, $H_M$ or $H_R$, which acts on the left end, the main part with notches, or the right end of the magnetic wire, respectively. $H_L$ ($H_R$) can trigger the nucleation of a DW at the left (right) end of the magnetic wire. Thus, the propagation direction of a DW can be controlled by $H_L$ ($H_R$). The resistance was measured by using a four-point dc technique, and an external magnetic field ($H_{ext}$) was applied along the wire axis. All magnetoresistance (MR) measurements were performed at room temperature.

Figure 2(a) shows a typical resistance change of the trilayerd magnetic wire with asymmetric notches as a function of $H_{ext}$. Before the measurement, $H_{ext}$ of -200 Oe was applied in order to align the magnetizations of two NiFe layers of the trilayered wire in the same direction. Then, MR measurement was performed while $H_{ext}$ was swept up to +200 Oe. When $H_{ext}$ was negative, the resistance showed the smallest value due to the parallel magnetization alignment of two NiFe layers. Then the resistance suddenly increased and showed the largest value with $H_{ext}$ between 10 and 162 Oe. We have an evidence from a previous study on NiFe wires of various thickness that a thicker NiFe layer has a larger coercive force than thinner one.[6] Therefore, this resistance increase corresponds to the magnetization reversal of the NiFe(5 nm) layer, and an antiparallel magnetization alignment was realized while the resistance showed the largest value. At 162 Oe, the resistance abruptly decreased from the largest value to the smallest value, which corresponds to the overall magnetization reversal of the NiFe(20 nm) layer. Thus, the magnetic DW was not pinned by asymmetric notches during the magnetization reversals of NiFe(20 nm) layer because of its large nucleation field.

In order to nucleate of a DW in the NiFe(20 nm) layer at smaller $H_{ext}$ and to pin the DW at the notch, we utilized the method generating a pulsed local magnetic field at an end of the magnetic wire.[5] Figure 2(b) shows the result of the DW injection into the NiFe(20 nm) layer of the magnetic wire by $H_L$. The measurement procedure is the same as that of the measurement shown in Fig. 2(a), except applying $H_L$ when $H_{ext}$ = 60 Oe. The magnitude and the duration of the pulsed $H_L$ were 200 Oe and 100 ns, respectively. The resistance abruptly decreased after the application of $H_L$ and stayed at a value between the largest and the smallest values. The value of the resistance indicates that a DW injected from the left by $H_L$ was pinned at the first notch. By increasing $H_{ext}$ after the injection of the DW, the resistance abruptly decreased to the smallest value at 98 Oe, which indicates that the DW propagated to the right end of the wire through the asymmetric notches. Thus, the depinning field for the rightward propagation of the DW can be determined to be 98 Oe. On the other hand, the depinning field for the leftward propagation of the DW can be determined from the result shown in Fig. 2(c), which shows the MR measurement when the DW was injected from the right end of the wire by $H_R$. In this case, the depinning field was 54 Oe. Thus, the depinning field for the rightward propagation is much larger than that for the rightward propagation. This result indicates that the asymmetric notch in the magnetic wire works as an asymmetric pinning potential against the DW propagation.

As seen in Figs. 2(b) and 2(c), when $H_{ext}$ exceeds the pinning field of the first notch, the DW propagates to the end of the wire and cannot be pinned at the successive notches. This indicates that all notches are identical from the viewpoint of the pinning potential. Here, we demonstrate that a combination of $H_{ext}$ and pulsed $H_M$ can move the DW from one notch to the neighboring one. This is shown in Fig. 3. The horizontal and the vertical axes show the elapsed time from the measurement start and the resistance of the sample, respectively. In the beginning of the measurement, $H_{ext}$ of -200 Oe was applied, and then $H_{ext}$ was swept up to 30 Oe and kept at 30 Oe until the measurement finished. When $H_{ext}$ reached 30 Oe, the resistance showed the largest value and the magnetizations of two NiFe layers became antiparallel. By the application of the pulsed $H_R$, a DW was injected into the NiFe(20 nm) layer and it was pinned at the first notch, which results in the first abrupt decrease in resistance. Then, a pulsed $H_M$ was applied to the same direction as $H_{ext}$. The magnitude and the duration of the pulsed $H_M$ were 20 Oe and 50 ns, respectively. Each $H_M$ triggered an abrupt decrease in resistance. The value of each resistance change corresponded to the length between the notches. Thus, the DW moved notch by notch by the pulsed $H_M$. We also confirmed that the DW moved notch by notch to the rightward direction by the pulsed $H_M$ with the opposite polarity.

The interpretation why the depinning field depends on the propagation direction is the following. Since the DW energy is proportional to its area, the DW at wider position in the wire has larger energy. This change in the DW energy along the wire axis produces the pinning potential for the DW. The force to move the DW against the potential is given by the derivative of the DW energy with respect to the DW position, which is proportional to the slope of the notch. In our magnetic wire shown in Fig. 1, the slope of the notch is steeper for the rightward propagation, resulting in the larger depinning field for the rightward propagation.

In conclusion, pinning and depinning of a magnetic DW in a submicron magnetic wire with asymmetric notches were investigated by utilizing the GMR effect. The DW was nucleated and injected from an end of the magnetic wire by a pulsed local magnetic field. The depinning field of the DW from the notch depends on the propagation direction of the DW. This indicates that the asymmetric structures constructed into the magnetic wire work as asymmetric pinning potential against the DW propagation.

The present work was partly supported by the Ministry of Education, Culture, Sports, Science and Technology of Japan (MEXT) through the Grants-in-Aid for COE Research (10CE2004, 12CE2005) and MEXT Special Coordination Funds for Promoting Science and Technology (Nanospintronics Design and Realization, NDR) and by the 21st Century COE Program by Japan Society for the Promotion of Science.

**Figure captions**

**Figure 1.** Schematic illustration of a top view of the whole sample. The black part is a trilayered magnetic wire consisting of NiFe(5 nm)/Cu(20 nm)/NiFe(20 nm). The main body of the magnetic wire has four asymmetric notches. A flow of an electric current in the Cu wire crossing the magnetic wire generates a local magnetic field, which is used to nucleate a magnetic DW.

**Figure 2.** Resistance change of the trilayered magnetic wire with asymmetric notches as a function of the external magnetic field. The magnetic domain structures inferred from the resistance measurement are schematically shown. **(a)** Typical MR curve of the trilayered system. **(b)** The magnetic DW was injected into the NiFe(20 nm) layer from the left end of the magnetic wire by the pulsed $H_\mathrm{L}$. **(c)** The DW was injected into the NiFe(20 nm) layer from the right end of the magnetic wire by the pulsed $H_\mathrm{R}$.

**Figure 3.** Resistance change of the MR measurement of the trilayered magnetic wire with asymmetric notches as a function of the elapsed time. The magnetic domain structures of the NiFe(20 nm) layer inferred from the resistance change are schematically shown.

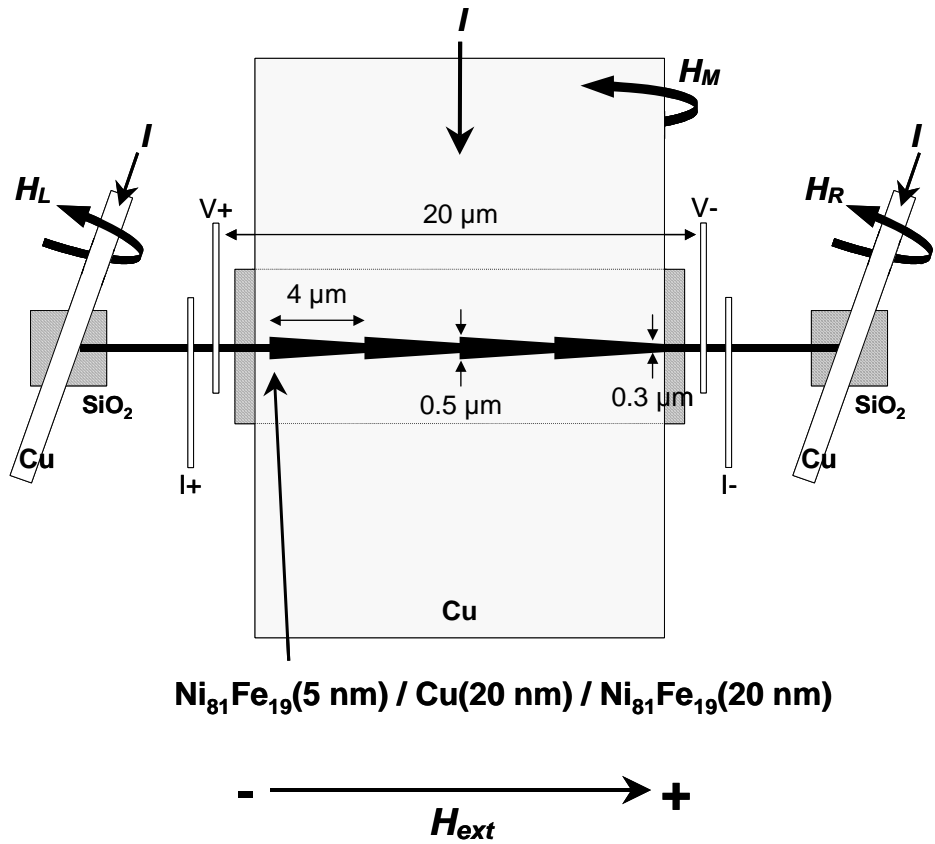

**Ni$_{81}$Fe$_{19}$(5 nm) / Cu(20 nm) / Ni$_{81}$Fe$_{19}$(20 nm)**

Fig. 1 - A. Himeno

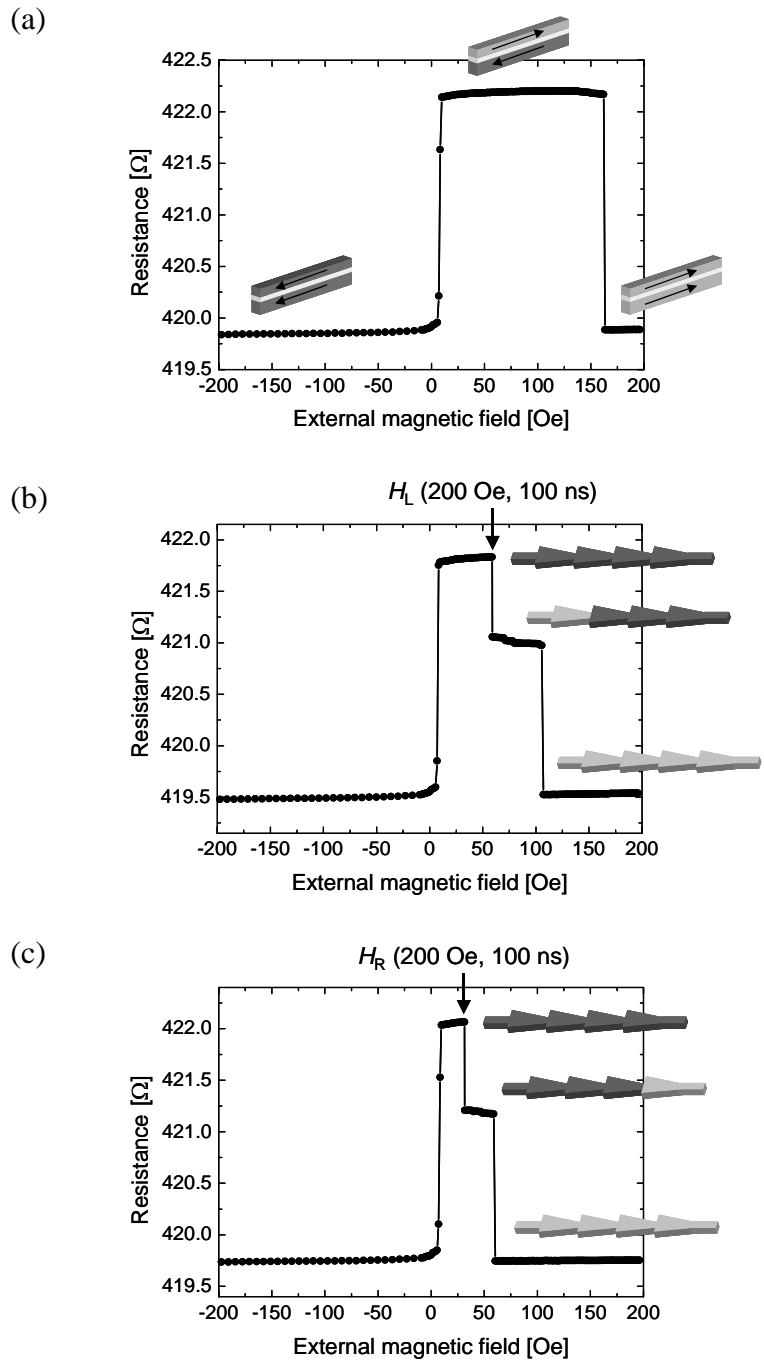

Fig. 2 - A. Himeno

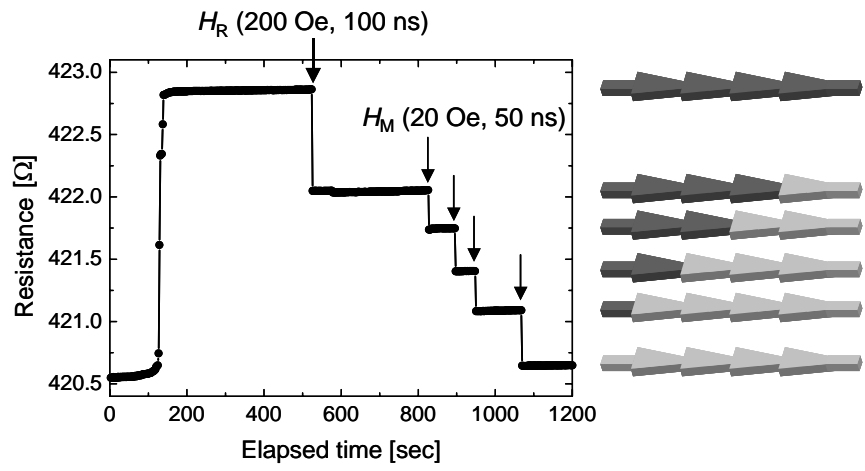

Fig. 3 - A. Himeno